\documentclass[a4paper,10pt]{article}

\usepackage{amsmath}            
\usepackage{amsfonts}           
\usepackage{amssymb}            

\begin{document}
\title{\huge \bf Universality of the Einstein theory of gravitation}

\author{Jerzy Kijowski
\\
Center for Theoretical Physics, Polish Academy of Sciences, \\
Al. Lotnik\'ow 32/46; 02-668 Warszawa, Poland
}

\date{}

\maketitle

\begin{abstract}
We show that generalizations of general relativity theory, which consist in replacing the Hilbert Lagrangian $L_{Hilbert} = \frac 1{16\pi} \sqrt{|g|} R$ by a generic scalar density $L=L(g_{\mu\nu}, R^\lambda_{\mu\nu\kappa})$ depending upon the metric $g_{\mu\nu}$ and the curvature tensor $R^\lambda_{\mu\nu\kappa}$, are equivalent to the conventional Einstein theory for a (possibly) different metric tensor $\tilde{g}_{\mu\nu}$ and (possibly) a different set of matter fields. The simple proof of this theorem relies on a new approach to variational problems containing metric and connection.
\end{abstract}

\section{Introduction}

Einstein theory of gravity can be derived from the variational principle:
\begin{equation}\label{Var}
    L(g, \partial g, \partial^2 g , \varphi, \partial \varphi ) = L_{Hilbert} (g, \partial g, \partial^2 g)
    + L_{Matter}(g, \partial g, \varphi, \partial \varphi ) \ ,
\end{equation}
with the universal Hilbert Lagrangian:
\begin{equation}\label{Hilb}
    L_{Hilbert} = \frac 1{16\pi} \sqrt{|g|} R = \pi^{\mu\nu} R_{\mu\nu}\ .
\end{equation}
Here, $R_{\mu\nu}$ is the Ricci tensor, $R$ is the scalar curvature. Moreover, by
\begin{equation}\label{pi-def}
    \pi^{\mu\nu} := \frac 1{16\pi} \sqrt{|g|} g^{\mu\nu} \ ,
\end{equation}
we have denoted the contravariant density of metric tensor.

Usually, one considers the so called ``minimal coupling'' of various matter fields (denoted symbolically by $\varphi$) with gravity. This requirement means that we begin with a special-relativistic version of the matter Lagrangian $L_{Matter}(g,  \varphi, \partial \varphi )$ and then we replace partial derivatives $\partial \varphi $ by appropriate ``covariant'' derivatives $\nabla \varphi = \partial \varphi +``\Gamma \cdot \varphi$'', where $\Gamma$ is the Levi-Civita connection. The last statement is merely symbolical. It makes a precise sense only for a tensor field $\varphi$. For an arbitrary matter field we only assume that $L_{Matter}$ is an invariant scalar density built from the field, the metric and their first derivatives.

But, we have:
\begin{equation}\label{var-Hilb}
    \frac {\delta L_{Hilbert}}{\delta g_{\mu\nu}} = - \frac 1{16\pi} \sqrt{|g|} G^{\mu\nu} \ ,
\end{equation}
where by $G$ we denote the Einstein tensor. Defining the ,,matter energy-momentum tensor'':
\begin{equation}\label{var-varphi}
    T^{\mu\nu} := \frac 2{\sqrt{|g|}} \frac {\delta L_{Matter}}{\delta g_{\mu\nu}}  \ .
\end{equation}
we obtain the following form of field equations of the theory:
\begin{eqnarray*}
  0  &=& \frac {\delta L}{\delta g_{\mu\nu}} = -  \frac 1{16\pi} \sqrt{|g|} \left( G^{\mu\nu} - 8\pi
  T^{\mu\nu}\right) \ , \\
  0  &=& \frac {\delta L}{\delta \varphi } =\frac {\delta L_{Matter}}{\delta \varphi } \ .
\end{eqnarray*}

Replacing Hilbert Lagrangian by an arbitrary scalar density $L$ depending upon $g$ and $R$, but no longer linear in the curvature, changes substantially the character of our theory. In a generic case, field equations are no longer of the second differential order with respect to the metric, but are fourth order PDE's.

Consider, therefore, a ,,generalized'' theory of gravity, based on an invariant Lagrangian:
\begin{equation}\label{inv}
    L = L(g_{\mu\nu}, R^\lambda_{\mu\nu\kappa} , \Gamma^\lambda_{\mu\nu}, \varphi, \partial\varphi) \ ,
\end{equation}
where $\Gamma^\lambda_{\mu\nu}$ is a Levi-Civita connection of the metric $g_{\mu\nu}$ and $R^\lambda_{\mu\nu\kappa}$ denotes its Riemann tensor. In this paper we prove the following mathematical statement.

\noindent
{\bf Theorem 1:} There exists a one-to-one change of variables:
\begin{equation}\label{change}
    (g , \varphi) \Longleftrightarrow (\tilde{g} , \varphi , \phi) \ ,
\end{equation}
and a new matter Lagrangian:
\begin{equation}\label{matter-new}
    \tilde{L}_{Matter} = \tilde{L}_{Matter}(\tilde{g}, \partial \tilde{g}, \varphi, \phi , \partial \varphi , \partial \phi) \ ,
\end{equation}
such that $(g,  \varphi)$ satisfy field equations derived from the Lagrangian \eqref{inv} if and only if the corresponding fields $(\tilde{g} , \varphi , \phi)$ satisfy the conventional ,,Einstein + matter'' equations, derived from the conventional variational principle:
\begin{equation}\label{L-Hilb-tilde}
    \tilde{L}:= L_{Hilbert}(\tilde{g}) + \tilde{L}_{Matter} \ .
\end{equation}
In particular, equations for the new metric $\tilde{g}$ are of the second differential order: $G^{\mu\nu}(\tilde{g}) = 8\pi \tilde{T}^{\mu\nu}$, where
\begin{equation}\label{var-varphi-phi}
    \tilde{T}^{\mu\nu} := \frac 2{\sqrt{|\tilde{g}|}} \frac {\delta \tilde{L}_{Matter}}{\delta \tilde{g}_{\mu\nu}}  \ .
\end{equation}
Also matter field equations are of the second differential order because $\tilde{L}_{Matter}$ depends upon first derivatives only.

To define new metric $\tilde{g}$ and new matter fields $\phi$ we decompose the curvature tensor into a sum of two irreducible components describing: 1) the Ricci tensor $R_{\mu\nu}$ and 2) the Weyl tensor $W^\lambda_{\mu\nu\kappa}$ (see Section \ref{An}, formulae \eqref{R-irreducible} or \eqref{K-irreducible}).
The new metric (or, rather, its contravariant density, cf. formula \eqref{pi-def})  is defined as the ``momentum canonically conjugate'' to the Ricci tensor:
\begin{equation}\label{tilde-pi-R}
       \tilde{\pi}^{\mu\nu} := \frac {\partial L}{\partial
  R_{\mu\nu}}\ ,
\end{equation}
whereas new matter fields $\phi$ describe: 1) the old metric $g$ and 2) the field $p_\lambda^{\mu\nu\kappa}$ defined as the ``momentum canonically conjugate'' to the Weyl tensor:
\begin{equation}\label{p-def-pierwsze}
    p_\lambda^{\mu\nu\kappa} := \frac {\partial L}{\partial
  W^\lambda_{\mu\nu\kappa}}\ .
\end{equation}
This means that the new fields $(\tilde{g}  , \phi)$ are defined as combinations of the old fields $(g , \varphi)$ and their derivatives up to the second order.
The new matter Lagrangian is calculated in Section \ref{proof} (see formula \eqref{L-matter-tilde}).

The particular case of a Lagrangian $L$ which depends non-linearly upon the Ricci tensor, but does not depend upon the Weyl tensor, was considered by many authors (cf.~\cite{Stelle}). Mathematical structure of such theories was thoroughly analyzed already long ago (see e.g.~\cite{Jakubiec}). In particular, equation \eqref{p-def-pierwsze} implies that the field $p_\lambda^{\mu\nu\kappa}$ vanishes identically. Hence, there is only one ``new matter field'', namely the old metric, arising in such models.
Probably the first, physically well motivated, proposal of such a theory was the Sacharov's non-linear Lagrangian containing the $R^2$ term (see \cite{Sacharov}). In this case, and also for any Lagrangian depending exclusively upon the scalar curvature $R$, i.e.~for $L=\sqrt{|g|}f(R)$, equation \eqref{tilde-pi-R} implies that the old metric $\pi$ is proportional to the new metric $\tilde{\pi}$. Indeed, \eqref{tilde-pi-R} reads:
\begin{equation}\label{Sacharov}
       \tilde{\pi}^{\mu\nu} := \frac {\partial L}{\partial
  R_{\mu\nu}} = f^\prime \sqrt{|g|} g^{\mu\nu} = \pi^{\mu\nu} e^{-\phi} \ .
\end{equation}
Consequently, the new matter field $\pi^{\mu\nu}$ can be encoded by a single scalar field $\phi$ (see \cite{Ferraris} and also \cite{Jakubiec}). Sacharov theory is, therefore, equivalent to the standard Einstein general relativity theory interacting with a non-linear scalar field\footnote{Also Brans-Dicke theory can be mentioned in this context. It is, however, much simpler to handle because its Lagrangian is linear in the curvature.}:
\begin{equation}\label{L-Hilb-tilde-Sacharov}
    \tilde{L}:= L_{Hilbert}(\tilde{g}) + \tilde{L}_{Matter} (\phi, \partial \phi, \tilde{g}) \ .
\end{equation}
Special examples of the Lagrangians depending upon the Ricci tensor: $L=L(g_{\mu\nu}, R_{\mu\nu})$, were analyzed also by Stephenson and Higgs (see \cite{Higgs}). These results are, however, purely algebraic and do not apply to a generic Lagrangian of this type. Moreover, the theories considered in \cite{Higgs} belong to a (much simpler) class of ``purely affine'' theories which we analyze in Section \ref{affine}. They differ considerably from the Einstein theory. In our paper we have rather in mind theories whose weak field limit do not differ substantially from General Relativity Theory.

Recently, there is a renewed interest in generalizations of Einstein theory of gravity (see e.g. \cite{FRA}). In this context our result can be summarized  as follows: generalizations of gravity theory based on non-conventional Lagrangians consist, practically, in introducing non-conventional matter fields, whose gravitational interaction, however, remains conventional (i.e. Einsteinian). We stress, that our result is mathematically rigorous, even if there might be doubts concerning physical interpretation of the change of variables \eqref{change}. Further discussion of physical aspects of this transformation is contained in Section 7, but we mainly concentrate on the mathematical structure of the theory.

The principal advantage of our result consists in the fact, that  various dynamical properties (i.g.~stability!) of generalized theories of gravity can be analyzed with help of the entire canonical (Hamiltonian) formalism of general relativity theory, including the positive mass theorem and its consequences. In particular, positivity of the total mass in generalized gravity is immediately equivalent with obvious energy conditions imposed on the new energy-momentum tensor \eqref{var-varphi-phi} of the new matter fields $(\tilde{g} , \varphi , \phi)$ (cf. also \cite{Jakubiec-Cauchy}).

Variational principles based on the curvature are, computationally, relatively complicated. Even standard textbooks, like Misner-Thorne-Wheeler (see e.g.~\cite{WMT}, formula 21.86 on page 520) try to avoid calculations and limit themselves to presentation of final results. In particular, boundary terms, which are necessary for the quasi-local description of gravitational energy (cf. \cite{CJK}) are never presented. However, they follow from the following, simple identity (see also \cite{pieszy}), which can easily be checked:
\begin{eqnarray}
    \delta L_{Hilbert}
    &=&
    - \frac 1{16 \pi} \sqrt{|g|} G^{\mu\nu} \delta g_{\mu\nu}
    + \partial_\kappa
    \left\{\left( \delta^\kappa_\lambda \pi^{\mu\nu} - \delta^{\mu}_\lambda \pi^{\nu\kappa}\right)
    \delta \Gamma^\lambda_{\mu\nu}
    \right\} \nonumber \\
    &=&
    R_{\mu\nu} \delta \pi^{\mu\nu} + \partial_\kappa
    \left\{\pi_\lambda^{\mu\nu\kappa}
    \delta \Gamma^\lambda_{\mu\nu}
    \right\}\ . \label{delt-Hilb}
\end{eqnarray}
The following tensor density with four indices:
\begin{equation}\label{duze-pi}
    \pi_\lambda^{\mu\nu\kappa} :=  \delta^\kappa_\lambda \pi^{\mu\nu} - \delta^{(\mu}_\lambda \pi^{\nu)\kappa}
    =\delta^\kappa_\lambda \pi^{\mu\nu} - \frac 12 \delta^{\mu}_\lambda \pi^{\nu\kappa} -
    \frac 12 \delta^{\nu}_\lambda \pi^{\mu\kappa} \ ,
\end{equation}
arises here in a natural way as a ``momentum'' canonically conjugate to the connection $\Gamma$. Formula \eqref{delt-Hilb} suggests to replace metric $g$ by its contravariant density \eqref{pi-def} in the variational principle. This method of variation was used by many authors since the classical times of General Relativity Theory
(see e.g.~the Fock's monograph \cite{Fock}, formula (60.14)). We use it also here to simplify the proof of our Theorem.
In particular, transition from the Hilbert to the so called Einstein Lagrangian (which is of the first differential order in $g$), presented usually as a transition between ``the sufficient part'' and the ``whole'' (see again Wheelr-Misner-Thorn \cite{WMT}, formula 21.85 on page 519), arises in this context as a simple Legendre transformation between connection and metric. We shortly sketch his formalism in Section~2 as an technical tool used in the proof of the Theorem.

Even more radical simplification is due to the novel mathematical description of the curvature, presented in Section 3. In Section 4 we combine this formalism with the variational principle and the enormous simplification of the theory becomes obvious. Section 5 contains a simple derivation of field equations in a ``generalized'' theory, based on variational principle \eqref{inv}. This Section is meant as an illustration of how our formalism works. Finally Section 6 contains proof of the Theorem and examples. Its possible implications are discussed in Section~7.

Recently, there is a growing interest in theories based on variational formulae containing higher order derivatives of metric (see \cite{Wald} and the references herein). The techniques developed in the present paper can be easily adapted to such theories. For this purpose, results obtained in \cite{Kij-Moreno} will be used and the corresponding results will be presented soon.

\section{From Hilbert to Einstein: a Legendre transformation}

The present Section constitutes, in fact, the concluding part of the proof of Theorem 1. We have shifted it here, because it illustrates the notation and the formalism used in this paper.

Equation \eqref{delt-Hilb} can be rewritten as follows:
\begin{equation}\label{delt-Hilb-2}
    \delta L_{Hilbert} =
    R_{\mu\nu} \delta \pi^{\mu\nu} + \left(\partial_\kappa \pi_\lambda^{\mu\nu\kappa}\right) \delta \Gamma^\lambda_{\mu\nu} +
    \pi_\lambda^{\mu\nu\kappa} \delta \Gamma^\lambda_{\mu\nu\kappa}\ ,
\end{equation}
where we use the ``jet adapted'' notation:
\begin{equation}\label{jet-Gamma}
    \Gamma^\lambda_{\mu\nu\kappa} := \partial_\kappa \Gamma^\lambda_{\mu\nu} \ ,
\end{equation}
and, whence, we have:
\[
    \pi_\lambda^{\mu\nu\kappa} = \frac {\partial L_{Hilbert}}{\partial \Gamma^\lambda_{\mu\nu\kappa}} \ .
\]
Similarly, we can introduce momenta canonically conjugate to the matter variables $\varphi$:
\[
    p^{\kappa} = \frac {\partial L_{Matter}}{\partial \varphi_\kappa} = \frac {\partial L}{\partial \varphi_\kappa} \ ,
\]
(again the ``jet adapted convention'': $\varphi_\kappa := \partial_\kappa \varphi$ is used).
Using the same convention for the metric (in its ``contavariant density'' representation):
\[
    S^{\kappa}_{\mu\nu} = \frac {\partial L_{Matter}}{\partial \pi^{\mu\nu}_{\ \ \kappa}}
     \ \ \ , \ \ \pi^{\mu\nu}_{\ \ \kappa} := \partial_\kappa \pi^{\mu\nu} \ ,
\]
we can finally rewrite the variation of the total Lagrangian \eqref{Var}:
\begin{eqnarray}
  \delta L &=& R_{\mu\nu} \delta \pi^{\mu\nu} + \partial_\kappa \left(\pi_\lambda^{\mu\nu\kappa} \delta \Gamma^\lambda_{\mu\nu}\right) + \frac {\partial L_{Matter}}{\partial \pi^{\mu\nu}} \delta \pi^{\mu\nu} +S^{\kappa}_{\mu\nu} \delta \pi^{\mu\nu}_{\ \ \kappa} \nonumber \\
  &+& \frac {\partial L_{Matter}}{\partial \varphi} \delta \varphi + p^\kappa \delta \varphi_\kappa \nonumber \\
    &=&  \left(R_{\mu\nu} + \frac {\partial L_{Matter}}{\partial \pi^{\mu\nu}} - \partial_\kappa S^{\kappa}_{\mu\nu}
    \right) \delta \pi^{\mu\nu} + \left( \frac {\partial L_{Matter}}{\partial \varphi}
    - \partial_\kappa p^\kappa \right) \delta\varphi \nonumber \\
    &+& \partial_\kappa \left(\pi_\lambda^{\mu\nu\kappa} \delta \Gamma^\lambda_{\mu\nu} +
    S^{\kappa}_{\mu\nu} \delta \pi^{\mu\nu} + p^\kappa \delta \varphi\right) \ . \label{EL-1}
\end{eqnarray}
Vanishing of the volume part of $\delta L$ is equivalent to the Euler-Lagrange equations of the theory. The equations consist, therefore, of two parts: 1) Einstein equations
\begin{equation}\label{Einst-cone}
    0 = \frac {\delta L}{\delta \pi^{\mu\nu}} = R_{\mu\nu} + \frac {\partial L_{Matter}}{\partial \pi^{\mu\nu}} - \partial_\kappa S^{\kappa}_{\mu\nu} \ ,
\end{equation}
and the matter field equations:
\begin{equation}\label{Einst-field-phi}
    0 = \frac {\delta L}{\delta \varphi} = \frac {\delta L_{Matter}}{\delta \varphi} = \frac {\partial L_{Matter}}{\partial \varphi}
    - \partial_\kappa p^\kappa \ .
\end{equation}
Vanishing of the volume part of $\delta L$ is, therefore, equivalent to the fact, that the variation reduces to its boundary part. Hence,
identity \eqref{EL-1} implies, that field equations \eqref{Einst-cone} -- \eqref{Einst-field-phi} can be rewritten in an equivalent way:
\begin{equation}\label{first-symp}
    \delta L = \partial_\kappa \left(\pi_\lambda^{\mu\nu\kappa} \delta \Gamma^\lambda_{\mu\nu} +
    S^{\kappa}_{\mu\nu} \delta \pi^{\mu\nu} + p^\kappa \delta \varphi\right) \ .
\end{equation}
To merely simplify our notation we denote
\begin{equation}\label{A-defi}
    A^\lambda_{\mu\nu} :=  \Gamma^\lambda_{\mu\nu} - \delta^\lambda_{(\mu} \Gamma^\alpha_{\nu)\alpha} = \Gamma^\lambda_{\mu\nu} - \frac 12 \delta^\lambda_{\mu} \Gamma^\alpha_{\nu\alpha} -
    \frac 12 \delta^\lambda_{\nu} \Gamma^\alpha_{\mu\alpha} \ ,
\end{equation}
and obtain:
\begin{equation}\label{pidGamma}
    \pi_\lambda^{\mu\nu\kappa} \delta \Gamma^\lambda_{\mu\nu}  =
    \pi^{\mu\nu} \delta A^\kappa_{\mu\nu} \ .
\end{equation}
No specific ``geometric'' interpretation has to be attached to $A$: it is merely a combination of components of $\Gamma$ which arises often in the sequel. Every formula containing $A$ has to be understood as a statement about the connection $\Gamma$! For example, Ricci tensor can be written shortly in terms of $A$:
\begin{equation}\label{R-w-A}
    R_{(\mu\nu)}= \partial_\lambda A^{\lambda}_{\mu\nu}
    -
A^{\lambda}_{\mu\sigma} A^{\sigma}_{\nu\lambda} + \frac 13
A^{\lambda}_{\mu\lambda} A^{\sigma}_{\nu\sigma} \ .
\end{equation}
The symmetrization is unnecessary in case of a metric connection. However, written as above, the formula is valid for any symmetric connection $\Gamma$.

At this point the following Legendre transformation between the connection and the metric can be performed:
\begin{equation}\label{Legendre}
    \pi_\lambda^{\mu\nu\kappa} \delta \Gamma^\lambda_{\mu\nu}  =
    \pi^{\mu\nu} \delta A^\kappa_{\mu\nu} = \delta \left(
    \pi^{\mu\nu} A^\kappa_{\mu\nu}\right) - A^\kappa_{\mu\nu} \delta \pi^{\mu\nu} \ .
\end{equation}
Consequently, field equations \eqref{first-symp} can be rewritten in yet another way:
\begin{equation}\label{first-symp-2}
    \delta \Lambda = \partial_\kappa \left\{ \left(- A^\kappa_{\mu\nu}  +
    S^{\kappa}_{\mu\nu}\right) \delta \pi^{\mu\nu} + p^\kappa \delta \varphi\right\} \ ,
\end{equation}
where the total derivative on the right hand side of \eqref{Legendre} has been shifted to the left hand side, producing the new Lagrangian $\Lambda$:
\begin{equation}\label{Lambda-def}
    \Lambda := L - \partial_\kappa \left( \pi^{\mu\nu}  A^\kappa_{\mu\nu}
    \right) =
    L_{Hilbert} - \partial_\kappa \left( \pi^{\mu\nu}  A^\kappa_{\mu\nu}
    \right) + L_{Matter} \ .
\end{equation}
The first two components sum up to the so called Einstein Lagrangian. Due to \eqref{R-w-A}, we have:
\begin{eqnarray}
  L_{Einstein} &:=& L_{Hilbert} - \partial_\kappa \left( \pi^{\mu\nu}  A^\kappa_{\mu\nu}
    \right) \label{divergence}\\
    &=&  \pi^{\mu\nu} \left(-
    A^{\lambda}_{\mu\sigma} A^{\sigma}_{\nu\lambda} + \frac 13
    A^{\lambda}_{\mu\lambda} A^{\sigma}_{\nu\sigma}\right) -  \pi^{\mu\nu}_{\ \ \lambda}
    A^{\lambda}_{\mu\nu} \nonumber \\
    &=& {\pi}^{\mu\nu} \left(
{\Gamma}^{\lambda}_{\mu\sigma} {\Gamma}^{\sigma}_{\nu\lambda} -
{\Gamma}^{\lambda}_{\mu\nu} {\Gamma}_{\lambda\sigma}^\sigma
    \right)  \ . \label{Eins-explicit}
\end{eqnarray}
The final form of $L_{Einstein}$ has been obtained by expressing both $A^{\lambda}_{\mu\nu}$ and the derivatives $\pi^{\mu\nu}_{\ \ \lambda}$  of the metric in terms of the Christoffel symbols $\Gamma$. Observe that $L_{Einstein}$ does not depend upon second derivatives of the metric.

Subtracting the complete divergence from the Hilbert Lagrangian, like in  formula \eqref{divergence}, is a standard trick used in general relativity theory (cf.~\cite{WMT}). Nevertheless, its interpretation in terms of a Legendre transformation \eqref{Legendre} is extremely useful. It clarifies considerably the canonical structure of the general relativity theory (cf. \cite{affine}, \cite{pieszy}) and, as will be seen in the sequel, almost trivializes the proof of our Theorem.

Denoting:
\begin{equation}\label{A-tylda}
    - A^\kappa_{\mu\nu}  +   S^{\kappa}_{\mu\nu} =: -a^\kappa_{\mu\nu}
\end{equation}
we can write field equations \eqref{first-symp-2} of the theory in the following way:
\begin{eqnarray}
    & &\delta \Lambda(\pi^{\mu\nu} , \pi^{\mu\nu}_{\ \ \kappa} , \varphi , \varphi_\kappa) = \partial_\kappa \left\{ - a^\kappa_{\mu\nu} \delta \pi^{\mu\nu} + p^\kappa \delta \varphi\right\} \nonumber \\
    &=&  - \left(\partial_\kappa  a^\kappa_{\mu\nu}\right) \delta \pi^{\mu\nu}
    - a^\kappa_{\mu\nu} \delta \pi^{\mu\nu}_{\ \ \kappa}
    + \left( \partial_\kappa p^\kappa\right) \delta \varphi +
     p^\kappa \delta \varphi_\kappa \ ,\label{sympletcic-a}
\end{eqnarray}
or, equivalently:
\begin{eqnarray*}
  - \partial_\kappa  a^\kappa_{\mu\nu} &=& \frac {\partial \Lambda}{\partial
  \pi^{\mu\nu}}\ , \\
  - a^\kappa_{\mu\nu} &=& \frac {\partial \Lambda}{\partial \pi^{\mu\nu}_{\ \ \kappa}} \ , \\
  \partial_\kappa p^\kappa &=& \frac {\partial \Lambda}{\partial \varphi} \ , \\
  p^\kappa &=& \frac {\partial \Lambda}{\partial \varphi_\kappa} \ .
\end{eqnarray*}
These are Euler-Lagrange equations of $\Lambda$.  Knowing metric tensor $\pi^{\mu\nu}$ we know also $\Gamma^\kappa_{\mu\nu}$ and, therefore, also $A^\kappa_{\mu\nu}$. Hence, we can reconstruct the matter tensor $S^{\kappa}_{\mu\nu}$ (and, consequently, its energy-momentum tensor) from \eqref{A-tylda}:
\begin{equation}\label{A-tylda-minus-A}
    S^{\kappa}_{\mu\nu} =: A^\kappa_{\mu\nu} -a^\kappa_{\mu\nu} \ .
\end{equation}
Also, the matter Lagrangian $L_{Matter}$ can be reconstructed from $\Lambda$, due to formulae \eqref{Lambda-def}, \eqref{divergence} and \eqref{Eins-explicit}, namely
\begin{eqnarray}
    L_{Matter} &=& \Lambda -
    L_{Einstein} =  \Lambda - {\pi}^{\mu\nu} \left(
{\Gamma}^{\lambda}_{\mu\sigma} {\Gamma}^{\sigma}_{\nu\lambda} -
{\Gamma}^{\lambda}_{\mu\nu} {\Gamma}_{\lambda\sigma}^\sigma
    \right) \ ,\label{L-from-Lambda}
\end{eqnarray}
where connection coefficients $\Gamma$ have to be expressed in terms of the metric $\pi$ and its derivatives.

Observe, that no assumptions about the algebraic structure of $\Lambda$ was necessary. An arbitrary function $\Lambda$ of the metric $\pi$, matter fields $\varphi$ and their first derivatives can be used as a starting point. We have just proved that the resulting theory, derived from a generic $\Lambda$ {\em via} eq.~\eqref{sympletcic-a}, will be precisely the general relativity theory in its Einstein version. As will become clear in Section \ref{proof}, this observation constitutes the concluding part of the proof of our Theorem (see also formula \eqref{varia-P-pi}).

All the formulae used above are absolutely classical. But there is an extremely powerful mathematical structure hidden here. In fact, formula \eqref{sympletcic-a} contains a natural (canonical) symplectic structure
\begin{eqnarray}\label{sympletcic-b}
    \omega &=& \partial_\kappa \left\{ - \delta a^\kappa_{\mu\nu} \wedge \delta \pi^{\mu\nu} + \delta p^\kappa
    \wedge \delta \varphi\right\} \nonumber \\
    &=&  - \delta\left(\partial_\kappa  a^\kappa_{\mu\nu}\right)\wedge \delta \pi^{\mu\nu}
    - \delta a^\kappa_{\mu\nu} \wedge \delta \pi^{\mu\nu}_{\ \ \kappa}
    + \delta \left( \partial_\kappa p^\kappa\right) \wedge \delta \varphi +
     \delta p^\kappa \wedge \delta \varphi_\kappa \ ,
\end{eqnarray}
existing in the space of first jets of sections of the bundle describing both the metric field and the matter fields (cf. \cite{Tulcz-Kij}, \cite{pieszy}, \cite{Kij-Moreno}). This is the source of the Hamiltonian structure of general relativity theory. The proof of our Theorem, which we give in this paper, can be viewed as a simple application of this mathematical structure.

\section{An alternative description of the curvature tensor} \label{An}

Variation with respect to a connection field $\Gamma$ leads to an ugly algebra, which obscures considerably description of the corresponding canonical (Hamiltonian) structure of the theory. This is due to the fact that the momentum canonically conjugate to $\Gamma$:
\begin{equation}\label{P-grav}
    P_\lambda^{\mu\nu\kappa} := \frac {\partial L}{\partial \Gamma^\lambda_{\mu\nu\kappa}}
\end{equation}
(where $\Gamma^\lambda_{\mu\nu\kappa} = \partial_\kappa \Gamma^\lambda_{\mu\nu}$, cf.~\eqref{jet-Gamma}), and the derivative of the Lagrangian with respect to the Riemann tensor:
\begin{equation}\label{R-grav}
    Q_\lambda^{\mu\nu\kappa} := \frac {\partial L}{\partial R^\lambda_{\mu\nu\kappa}} \ ,
\end{equation}
although related by  a one-to-one correspondence, have different symmetries (symmetry {\em versus} antisymmetry; cf.~also \cite{WMT}, formula (21.20) on p.~500). Also Bianchi I-st type identities are implemented in a completely different way on $P$ and $Q$. Below, we propose an alternative description of the curvature, which trivializes this relation and, as will be seen in the sequel, simplifies enormously the proof of our Theorem.

Define a ``reference frame at a point ${\bf x} \in M$'' of a manifold $M$ as an equivalence class of coordinate charts with respect to the following relation ``$\sim_{\bf x}$''. Given two charts in a neighbourhood of ${\bf x}$, we declare them to be equivalent if the second derivatives of any coordinate from one chart with respect to coordinates of the other chart vanish at ${\bf x}$:
\begin{equation}\label{equiv-charts}
    \left( (x^\mu) \sim_{\bf x} (y^\alpha) \right) \Longleftrightarrow
    \left( \frac {\partial^2 y^\alpha}{\partial x^\mu x^\nu}({\bf x}) = 0
    \right) \ .
\end{equation}
It is easy to check that, indeed, it is an equivalence relation.

Given a reference frame $\Upsilon_0$ at ${\bf x}$, we may parameterize any other reference frame $\Upsilon$ at ${\bf x}$ by the following table of numbers:
\begin{equation}\label{Gamma-def}
    \Gamma^\lambda_{\mu\nu}({\bf x}) := \frac {\partial x^\lambda}{\partial y^\alpha}
    \frac {\partial^2 y^\alpha}{\partial x^\mu x^\nu}({\bf x}) \ ,
\end{equation}
where $(y^\alpha)$ is a representative of $\Upsilon_0$ and $(x^\mu)$ a representative of $\Upsilon$. It is easy to check that $\Gamma^\lambda_{\mu\nu}({\bf x})$ does not depend upon the choice of these representatives. This way the set of all reference frames acquires a structure of an affine fiber bundle over $M$.

Connection on a manifold $M$ is a ``field of reference frames'' $M\ni {\bf x} \rightarrow \Upsilon ({\bf x})$, i.e. a section of this bundle. The ``privileged'' reference frame $\Upsilon ({\bf x})$ at ${\bf x} \in M$ can be called a ``{\em local inertial frame} at ${\bf x}$''. Its coordinate description with respect to any coordinate chart $(x^\mu)$ is provided by the set of functions $\Gamma^\lambda_{\mu\nu}=\Gamma^\lambda_{\mu\nu}({\bf x})$. If $(x^\mu)$ belongs to this privileged class:  $(x^\mu) \in \Upsilon({\bf x})$, i.e.~if $\Gamma^\lambda_{\mu\nu}({\bf x})=0$, then $x^\mu$ will be called ``inertial coordinates at ${\bf x}$''.

Connection is {\em flat} if there exists a {\em global} inertial frame, i.e. a coordinate chart which is inertial not just at a single point, but everywhere. Given a connection $\Upsilon$, how to check whether or not it is flat? First, we can choose coordinates $(x^\mu)$ which are inertial at ${\bf x}$, i.e.~such that $\Gamma^\lambda_{\mu\nu}$ vanish at ${\bf x}$. Without any loss of generality we can assume that ${\bf x}=(0,0,\dots , 0)$. Is it possible to ``improve'' these coordinates in such a way that $\Gamma^\lambda_{\mu\nu}$ vanish also outside of ${\bf x}$? As a first step to answer this question let us try to kill also the derivatives $\Gamma^\lambda_{\mu\nu\kappa}({\bf x})= \partial_\kappa \Gamma^\lambda_{\mu\nu}({\bf x})$. Is it possible?

Consider such an improved system of coordinates:
\begin{equation}\label{improved}
    y^\lambda := x^\lambda + \frac 16 Q^\lambda_{\mu\nu\kappa} x^\mu x^\nu x^\kappa  + {\rm term \ of \ order \ higher \ than \ 3} \ ,
\end{equation}
where coefficients $Q$ are symmetric: $Q^\lambda_{\mu\nu\kappa} = Q^\lambda_{(\mu\nu\kappa)}$.
Only such coordinate transformations are interesting because:
\begin{enumerate}
  \item terms of order $0$ vanish under differentiation \eqref{Gamma-def}, i.e.~do not influence the connection coefficients $\Gamma^\lambda_{\mu\nu}$;
  \item terms of order $1$ produce only a linear (with constant coefficients) transformation of $\Gamma^\lambda_{\mu\nu}$ and, whence, a linear homogeneous (tensorial type) transformation of the coefficients $\Gamma^\lambda_{\mu\nu\kappa}({\bf x})$: if they do not vanish before, they will not vanish after such a transformation;
  \item non-vanishing terms of order $2$ would change, due to \eqref{Gamma-def}, the value of $\Gamma$ at ${\bf x}$. We try to avoid it because we have already $\Gamma^\lambda_{\mu\nu}({\bf x})=0$ and we do not want to spoil this!
  \item a possible non-symmetric part of $Q$ vanishes when contracted with the totally symmetric expression $x^\mu x^\nu x^\kappa$;
  \item 4th and higher order terms produce 2nd and higher order term in $\Gamma^\lambda_{\mu\nu}$ and, whence, do not change the value of derivatives $\Gamma^\lambda_{\mu\nu\kappa}({\bf x})$.
\end{enumerate}
Using \eqref{Gamma-def} we calculate the new connection coefficients $\tilde{\Gamma}$. They contain an extra linear term proportional to $Q$. Finally, after differentiation, we obtain:
\begin{equation}\label{Gamma-impro}
    \tilde{\Gamma}^\lambda_{\mu\nu\kappa}({\bf x}) = \Gamma^\lambda_{\mu\nu\kappa}({\bf x}) + Q^\lambda_{\mu\nu\kappa} \ .
\end{equation}
Using an arbitrary (but symmetric!) tensor $Q^\lambda_{\mu\nu\kappa}$ we are able to kill the totally symmetric part $\Gamma^\lambda_{(\mu\nu\kappa)}$ of $\Gamma^\lambda_{\mu\nu\kappa}$. The remaining part, if any:
\begin{equation}\label{K-def}
    K^\lambda_{\mu\nu\kappa} := \Gamma^\lambda_{\mu\nu\kappa} -
    \Gamma^\lambda_{(\mu\nu\kappa)} \ ,
\end{equation}
constitutes an obstruction against a possibility of killing derivatives of $\Gamma$, i.e. against its flatness. It measures, therefore, how {\em non-flat}, i.e.~how {\em curved}, is the connection. We call it the {\em curvature tensor}.

The above formula is valid in inertial coordinates. In generic coordinate system we calculate the value of the curvature tensor \eqref{K-def} at a point ${\bf x}$ in three steps: 1) recalculate $\Gamma$ to any inertial frame at ${\bf x}$, 2) calculate curvature tensor $K$ according to \eqref{K-def} and, finally: 3) recalculate components of the tensor $K$ back to original coordinate system. It is easy to prove that this way we obtain the following, universal formula, valid in an arbitrary coordinate system:
\begin{eqnarray}
    K^\lambda_{\mu\nu\kappa} &=&
    \Gamma^\lambda_{\mu\nu\kappa} - \Gamma^\lambda_{(\mu\nu\kappa)} + \left( \Gamma^{\lambda}_{\sigma\kappa}
  \Gamma^{\sigma}_{\mu\nu} - \Gamma^{\lambda}_{\sigma (\kappa} \Gamma^{\sigma}_{\mu\nu)}\right) \nonumber \\
  &=& \Gamma^\lambda_{\mu\nu\kappa} + \Gamma^{\lambda}_{\sigma\kappa}
  \Gamma^{\sigma}_{\mu\nu} - \left(\Gamma^\lambda_{(\mu\nu\kappa)} + \Gamma^{\lambda}_{\sigma(\kappa} \Gamma^{\sigma}_{\mu\nu)}\right) \ . \label{Krzywizna-dow}
\end{eqnarray}
Due to the definition, the curvature tensor $K$ is symmetric in first indices and its totally symmetric part vanishes:
\begin{equation}\label{K-symmetries}
    K^\lambda_{\mu\nu\kappa} = K^\lambda_{\nu\mu\kappa} \ \ \ ; \ \ \
    K^\lambda_{(\mu\nu\kappa)} = 0\ .
\end{equation}
The last identity can be called Bianchi I-st type identity.

The above curvature tensor is equivalent to the standard Riemann tensor $R^\lambda_{\mu\nu\kappa}$: antisymmetrization of $K$ in last two indices produces $R$ and symmetrization of $R$ in first two indices produces $K$. More precisely, the following relations are obvious:
\begin{equation}\label{rowno-KR}
    R^\lambda_{\mu\nu\kappa} = -2 K^\lambda_{\mu[\nu\kappa]} \ \ \  ; \ \ \ \ \
    K^\lambda_{\mu\nu\kappa} = - \frac{2}{3}R^\lambda_{(\mu\nu)\kappa} \ ,
\end{equation}
and the identities \eqref{K-symmetries} for $K$ are equivalent to the analogous identities for $R$:
\begin{equation}\label{R-symmetries}
    R^\lambda_{\mu\nu\kappa} = - R^\lambda_{\mu\kappa\nu} \ \ \ ; \ \ \
    R^\lambda_{[\mu\nu\kappa]} = 0\ .
\end{equation}

The curvature tensor can be decomposed into three irreducible parts: the symmetric and antisymmetric part of the Ricci tensor and the traceless (Weyl)  tensor. More precisely, we have:
\begin{equation}\label{K-irreducible}
    K^\lambda_{\mu\nu\kappa} = - \frac 19 \left( \delta^\lambda_\mu K_{\nu\kappa} +
    \delta^\lambda_\nu K_{\mu\kappa} - 2 \delta^\lambda_\kappa K_{\mu\nu} \right)
    - \frac 15 \left( \delta^\lambda_\mu F_{\nu\kappa} +
    \delta^\lambda_\nu F_{\nu\kappa}  \right) + U^\lambda_{\mu\nu\kappa} \ ,
\end{equation}
where $K_{\mu\nu} = K_{\nu\mu}$ and $F_{\mu\nu} = -F_{\nu\mu}$. All the three terms on the right hand side of \eqref{K-irreducible} satisfy the same symmetries \eqref{K-symmetries}. Moreover, the last term is traceless: $U^\lambda_{\mu\nu\lambda}=U^\lambda_{\lambda\nu\kappa}=0$. The coefficients have been chosen in such a way that $K_{\mu\nu}$ and $F_{\mu\nu}$ are respectively the symmetric and the antisymmetric part of the Ricci tensor:
\begin{equation}\label{Ricci}
    R_{\mu\nu} := R^\lambda_{\mu\lambda\nu} = K_{\mu\nu} + F_{\mu\nu} \ .
\end{equation}
The traces of the curvature $K^\lambda_{\mu\nu\kappa}$ can be obtained from \eqref{K-irreducible}:
\begin{equation}\label{slady-K}
    K^\lambda_{\lambda\nu\kappa} = - \frac 13 K_{\nu\kappa} - F_{\nu\kappa} \ \ \ ; \ \ \   K^\lambda_{\mu\nu\lambda} = \frac 23 K_{\mu\nu} \ .
\end{equation}
For the sake of completeness let us mention that the corresponding  decomposition of the Riemann tensor can be obtained directly from \eqref{K-irreducible} and \eqref{rowno-KR}:
\begin{equation}\label{R-irreducible}
    R^{\lambda}_{\mu \nu \kappa } = \frac 13 \left( \delta^\lambda_\nu  K_{\mu \kappa } -
    \delta^\lambda_\kappa  K_{\mu \nu }  \right) + \frac 15
    \left( 2 \delta^\lambda_\mu  F_{\nu \kappa } +
    \delta^\lambda_\nu  F_{\mu \kappa } -  \delta^\lambda_\kappa  F_{\mu \nu } \right) + W^{\lambda}_{\mu \nu \kappa } \ ,
\end{equation}
where the Weyl tensor $W$ fulfills identities  \eqref{R-symmetries} and is traceless.
We have also:
\begin{equation}\label{Ricci-F}
    R^\lambda_{\lambda\mu\nu} = 2 F_{\mu\nu} \ .
\end{equation}
If $\Gamma$ is a metric connection, the second part of both \eqref{K-irreducible} and \eqref{R-irreducible} vanishes because we have $F_{\mu\nu}=0$ in this case. Finally, observe that $U$ contains the complete information about the Weyl tensor $W$ because we have:
\begin{equation}\label{rowno-UW}
    W^\lambda_{\mu\nu\kappa} = -2 U^\lambda_{\mu[\nu\kappa]} \ \ \  ; \ \ \ \ \
    U^\lambda_{\mu\nu\kappa} = - \frac{2}{3}W^\lambda_{(\mu\nu)\kappa} \ .
\end{equation}

\section{Affine variational principle. Field equations}\label{affine}

To prepare the techniques which are necessary for the purposes of our Theorem, we consider in this Section a (much simpler) ``purely affine'' variational principle. It covers, in particular, examples considered by Higgs including his ``unsolved case 3'' (see \cite{Higgs}). At the end of this Section we show how to solve this unsolved case.
Assume, therefore, that $L$ is a scalar density depending upon a curvature tensor $K^\lambda_{\mu\nu\kappa}$. The formula
\begin{equation}\label{der-sym}
    \delta L(K^\lambda_{\mu\nu\kappa}) = P_\lambda^{\mu\nu\kappa} \delta K^\lambda_{\mu\nu\kappa} \ ,
\end{equation}
does not define uniquely its derivative
\begin{equation}\label{der-sym1}
    P_\lambda^{\mu\nu\kappa} = \frac {\partial L}{\partial K^\lambda_{\mu\nu\kappa}} \ ,
\end{equation}
unless we assume symmetries of $P$ dual to symmetries \eqref{K-symmetries} of the curvature tensor:
\begin{equation}\label{P-symmetries}
    P_\lambda^{\mu\nu\kappa} = P_\lambda^{\nu\mu\kappa} \ \ \ ; \ \ \
    P_\lambda^{(\mu\nu\kappa)} = 0\ .
\end{equation}
Due to this condition, which will always be imposed in the sequel, derivative \eqref{der-sym1} of $L$ can be uniquely represented by the tensor density $P$.

We admit also the dependence of $L$ upon matter fields and a conncetion:
\begin{equation}\label{Lagr-aff}
    L=L(K^\lambda_{\mu\nu\kappa}, \Gamma^\lambda_{\mu\nu} , \varphi , \varphi_\kappa) \ ,
\end{equation}
where $\varphi$ represents matter fields and $\varphi_\kappa :=\partial_\kappa \varphi$. Connection coefficients $\Gamma$ are already contained in the curvature tensor $K$, but we admit that they enter into $L$ {\em via} hypothetic ``covariant derivatives'' of matter fields. Observe that (at the moment) there is {\em no} metric tensor here and, therefore, no {\em metricity condition} is imposed on $\Gamma$. Field equations obtained from variation with respect to $\Gamma$ and $\varphi$ can be written in the following way:
\begin{eqnarray}
    \delta L &=& \partial_\kappa \left( P_\lambda^{\mu\nu\kappa} \delta \Gamma^\lambda_{\mu\nu} + p^\kappa \delta \varphi\right) \nonumber \\
    &=& \left(\partial_\kappa  P_\lambda^{\mu\nu\kappa}\right) \delta \Gamma^\lambda_{\mu\nu}
    + P_\lambda^{\mu\nu\kappa} \delta \Gamma^\lambda_{\mu\nu\kappa }+ \left(\partial_\kappa p^\kappa\right) \delta \varphi + p^\kappa \delta \varphi_\kappa \ ,\label{varia-A}
\end{eqnarray}
where the second and the last terms on the right hand side contain definitions of the momenta $P_\lambda^{\mu\nu\kappa}$ and $p^\kappa$, whereas the first and the third terms are the Euler-Lagrange equations of the theory. To obtain the ``covariant'' form of these equations let us replace the partial derivative  $\partial_\kappa  P_\lambda^{\mu\nu\kappa}$ in the first term by the corresponding covariant derivative. For this purpose we use the following identity, which is proved in the Appendix:
\begin{eqnarray}
    \nabla_\kappa P_\lambda^{\mu\nu\kappa}
    &=& \partial_\kappa P_\lambda^{\mu\nu\kappa} -
    P_\sigma^{\mu\nu\kappa} \Gamma^\sigma_{\lambda\kappa}
    -  P_\lambda^{\sigma\kappa(\mu} \Gamma^{\nu)}_{\sigma\kappa}  \ . \label{nabla-pi-konc}
\end{eqnarray}
Hence, we have:
\begin{eqnarray}
    & &\left(\partial_\kappa  P_\lambda^{\mu\nu\kappa}\right) \delta \Gamma^\lambda_{\mu\nu}
    + P_\lambda^{\mu\nu\kappa} \delta \Gamma^\lambda_{\mu\nu\kappa } \nonumber \\
    &=&
    \left( \nabla_\kappa P_\lambda^{\mu\nu\kappa} +
    P_\sigma^{\mu\nu\kappa} \Gamma^\sigma_{\lambda\kappa}
    +  P_\lambda^{\sigma\kappa(\mu} \Gamma^{\nu)}_{\sigma\kappa} \right)
    \delta \Gamma^\lambda_{\mu\nu}
    + P_\lambda^{\mu\nu\kappa} \delta \Gamma^\lambda_{\mu\nu\kappa } \nonumber \\
    &=& \left( \nabla_\kappa P_\lambda^{\mu\nu\kappa} \right)
    \delta \Gamma^\lambda_{\mu\nu}
    +  P_\lambda^{\mu\nu\kappa} \left(  \Gamma^{\lambda}_{\sigma\kappa} \delta \Gamma^\sigma_{\mu\nu}
    + \Gamma^\sigma_{\mu\nu} \delta \Gamma^{\lambda}_{\sigma\kappa}
    +  \delta \Gamma^\lambda_{\mu\nu\kappa } \right)\nonumber \\
    &=& \left( \nabla_\kappa P_\lambda^{\mu\nu\kappa} \right)
    \delta \Gamma^\lambda_{\mu\nu}
    + P_\lambda^{\mu\nu\kappa} \delta K^\lambda_{\mu\nu\kappa } \ ,\label{trans-P}
\end{eqnarray}
the last equality being implied by definition \eqref{Krzywizna-dow} of the curvature tensor and the ``Bianchi-like'' symmetry \eqref{P-symmetries}. Hence, field equations \eqref{varia-A} can be rewritten as:
\begin{eqnarray}\label{varia-A-B}
    \delta L &=& \left(\nabla_\kappa  P_\lambda^{\mu\nu\kappa}\right) \delta \Gamma^\lambda_{\mu\nu}
    + P_\lambda^{\mu\nu\kappa} \delta K^\lambda_{\mu\nu\kappa }+ \left(\partial_\kappa p^\kappa\right) \delta \varphi + p^\kappa \delta \varphi_\kappa \ ,
\end{eqnarray}
or, equivalently:
\begin{eqnarray*}
  P_\lambda^{\mu\nu\kappa} &=& \frac {\partial L}{\partial
  K^\lambda_{\mu\nu\kappa }}\ , \\
  \nabla_\kappa  P_\lambda^{\mu\nu\kappa} &=& \frac {\partial L}{\partial \Gamma^\lambda_{\mu\nu}} \ , \\
  p^\kappa &=& \frac {\partial L}{\partial \varphi_\kappa} \ , \\
  \partial_\kappa p^\kappa &=& \frac {\partial L}{\partial \varphi}
   \ .
\end{eqnarray*}
In particular, if the connection coefficients $\Gamma$ enter into the Lagrangian \eqref{Lagr-aff} only {\em via} the curvature $K$ (i.e.~if there are no ``covariant'' derivatives of the matter fields, like for e.g.~the scalar and the electromagnetic field) then the second equation reads: $\nabla_\kappa  P_\lambda^{\mu\nu\kappa}=0$.

The main advantage of the use of $K^\lambda_{\mu\nu\kappa }$ instead of the Riemann tensor $R^\lambda_{\mu\nu\kappa }$ consists in the identity
\[
    \frac {\partial L}{\partial  \Gamma^\lambda_{\mu\nu\kappa }} =\frac {\partial L}{\partial   K^\lambda_{\mu\nu\kappa }} \ .
\]
This (mathematically very modest!) achievement makes the tedious variational formulae several times shorter than in the standard formalism. This is probably the reason why such a simple fact as our Theorem has been overlooked so far.

Using decomposition \eqref{K-irreducible} and identities \eqref{P-symmetries} we can also decompose the momentum $P$ into three irreducible pieces:
\begin{eqnarray}
    P_\lambda^{\mu\nu\kappa} \delta K^\lambda_{\mu\nu\kappa } &=& - \frac 19 P_\lambda^{\mu\nu\kappa} \delta \left( \delta^\lambda_\mu K_{\nu\kappa} +
    \delta^\lambda_\nu K_{\mu\kappa} - 2 \delta^\lambda_\kappa K_{\mu\nu} \right)  \nonumber \\
    & & - \frac 15 P_\lambda^{\mu\nu\kappa} \delta
      \left( \delta^\lambda_\mu F_{\nu\kappa} +
    \delta^\lambda_\nu F_{\mu\kappa}  \right) + P_\lambda^{\mu\nu\kappa} \delta U^\lambda_{\mu\nu\kappa} \nonumber \\
    &=& \tilde{\pi}^{\mu\nu} \delta K_{\mu\nu} + {\cal F}^{\mu\nu}\delta F_{\mu\nu} + p_\lambda^{\mu\nu\kappa}
    \delta U^\lambda_{\mu\nu\kappa} \ , \label{P-irreducible-0}
\end{eqnarray}
where
\begin{eqnarray*}
    \tilde{\pi}^{\mu\nu} &=&  \frac 13 P_\lambda^{\mu\nu\lambda}  \ , \\
    {\cal F}^{\mu\nu} &=& - \frac 25 P_\lambda^{\lambda[\mu\nu]} \ ,
\end{eqnarray*}
and $p_\lambda^{\mu\nu\kappa}$ is the tracelss part of $P_\lambda^{\mu\nu\kappa}$. The following decomposition of $P_\lambda^{\mu\nu\kappa}$, dual with respect to \eqref{K-irreducible}, can be easily proved:
\begin{equation}\label{P-irreducible}
    P_\lambda^{\mu\nu\kappa} =  \left(\delta_\lambda^\kappa \tilde{\pi}^{\mu\nu} - \delta_\lambda^{(\mu} \tilde{\pi}^{\nu)\kappa}  \right)
    - \frac 12 \left( \delta_\lambda^\mu {\cal F}^{\nu\kappa} +
    \delta_\lambda^\nu {\cal F}^{\mu\kappa}  \right) + p_\lambda^{\mu\nu\kappa} \ .
\end{equation}
We conclude that $\tilde{\pi}$, ${\cal F}$ and $p$ are equal to the corresponding derivatives of the Lagrangian $L$ with respect to $K$, $F$ and $U$, respectively. Observe that the first term of the decomposition \eqref{P-irreducible} is analogous to the tensor \eqref{duze-pi} in the purely metric theory. This is how the ``true'' metric arises in a purely affine theory (cf.~\cite{Tulcz-Kij}, \cite{affine}). We denote it by $\tilde{\pi}$ because, in a generic case, it differs from the original metric tensor denoted by $\pi$. As will be seen in the sequel, the latter will be downgraded to the level of matter fields.

\section{Metric-affine lagrangian: a simple way to derive field equations}

Finally, we are ready to analyze our ``generalized'' general relativity theory, based on the Lagrangian function \eqref{inv}. In the present Section we show how the symplectic formalism introduced above leads to a simple derivation of the field equations. The reader who is interested in the proof of our Theorem only, can simply skip this Section. Nevertheless, it provides a good exercise and shows how powerful is the symplectic formalism in field theory.

As we already know, a considerable simplification of the structure is obtained if the metric tensor is encoded by the tensor density \eqref{pi-def} and the Riemann tensor is replaced by the curvature tensor \eqref{Krzywizna-dow}. Hence, we consider the Lagrangian
\begin{equation}\label{inv-new}
    L = L(\pi^{\mu\nu}, K^\lambda_{\mu\nu\kappa}(\Gamma) , \Gamma^\lambda_{\mu\nu}, \varphi, \varphi_\kappa) \ ,
\end{equation}
where $\Gamma$ is the Levi-Civita connection of the metric $\pi$. This is a constraint which can be formulated in the following way:
\begin{equation}\label{constr-G}
    \nabla_\kappa \pi^{\mu\nu} = 0 \ .
\end{equation}
Without any loss of generality, we consider only symmetric connections, because a non-symmetric connection can always be decomposed into a symmetric connection and a torsion. The latter will always be treated as one of the matter fields $\varphi$. Field equations of the theory can be obtained from the generating formula similar to \eqref{varia-A}. There are, however, two differences which makes our job a little bit more difficult: 1) there is an extra ``matter field'' $\pi$ and, whence, an extra momentum is necessary; moreover: 2) the constraints \eqref{constr-G} must be satisfied. Hence the following generating formula is true:
\begin{eqnarray}
    \delta L &=& \partial_\kappa \left( P_\lambda^{\mu\nu\kappa} \delta \Gamma^\lambda_{\mu\nu} + b^\kappa_{\mu\nu} \delta \pi^{\mu\nu} + p^\kappa \delta \varphi\right) \ ,\label{varia-A-G}
\end{eqnarray}
but variations on the right hand side must obey constraint equations \eqref{constr-G}. Using  \eqref{trans-P} from the previous Section, we rewrite the generating formula as follows:
\begin{eqnarray}
    & &\delta L(\pi^{\mu\nu}, K^\lambda_{\mu\nu\kappa} , \Gamma^\lambda_{\mu\nu}, \varphi, \varphi_\kappa)
    = \left(\nabla_\kappa  P_\lambda^{\mu\nu\kappa}\right) \delta \Gamma^\lambda_{\mu\nu}
    + P_\lambda^{\mu\nu\kappa} \delta K^\lambda_{\mu\nu\kappa } \nonumber \\
    &+&
    \left(\partial_\kappa  b^\kappa_{\mu\nu}\right) \delta \pi^{\mu\nu} + b^\kappa_{\mu\nu} \delta \pi^{\mu\nu}_{\ \ \kappa}
    +
    \left(\partial_\kappa p^\kappa\right) \delta \varphi + p^\kappa \delta \varphi_\kappa \ .\label{varia-A-G-inv}
\end{eqnarray}
Contrary to the quantity \eqref{A-defi} which arises in formula \eqref{first-symp-2}, the components $b^\kappa_{\mu\nu}$ represent a tensor and not just a connection coefficients because, in contrast to $\Lambda$, our present $L$ is an invariant scalar density.
On the other hand, constraints \eqref{constr-G} can be rewritten as:
\begin{equation}\label{constr-G-rev}
    0=\nabla_\kappa \pi^{\mu\nu} = \pi^{\mu\nu}_{\ \ \kappa}
    + \pi^{\sigma\nu} \Gamma^\mu_{\sigma\kappa} + \pi^{\mu\sigma} \Gamma^\nu_{\sigma\kappa} - \pi^{\mu\nu} \Gamma^\sigma_{\kappa\sigma}\ .
\end{equation}
(The last term arises because $\pi$ is a tensor density!). Equality \eqref{varia-A-G-inv} must be satisfied up to a covector which vanishes on the constraint submanifold  (see also \cite{Kij-Moreno}), i.e.~up to a combination $\Lambda^\kappa_{\mu\nu} \delta \left( \nabla_\kappa \pi^{\mu\nu} \right)$, where $\Lambda^\kappa_{\mu\nu}$ are ``Lagrange multiplyers''. But \eqref{constr-G-rev} implies:
\begin{eqnarray}\label{delta-constr}
    \Lambda^\kappa_{\mu\nu} \delta \left( \nabla_\kappa \pi^{\mu\nu} \right) &=&
    \Lambda^\kappa_{\mu\nu} \delta \pi^{\mu\nu}_{\ \ \kappa} +
    \left( \Lambda^\mu_{\lambda\kappa} \pi^{\nu\kappa} +\Lambda^\nu_{\lambda\kappa} \pi^{\mu\kappa}
    - \delta^{(\mu} \Lambda^{\nu )}_{\alpha\beta}\pi^{\alpha\beta}\right)
    \delta \Gamma^\lambda_{\mu\nu}  \nonumber \\
    &+& \left( \Lambda^\kappa_{\lambda\mu} \Gamma^\lambda_{\nu\kappa} + \Lambda^\kappa_{\lambda\nu} \Gamma^\lambda_{\mu\kappa} - \Lambda^\kappa_{\mu\nu} \Gamma^\lambda_{\kappa\lambda}
    \right)   \delta \pi^{\mu\nu}  \ .
\end{eqnarray}
Putting this expression on the right hand side of \eqref{varia-A-G-inv} and observing that $L$ does not contain derivatives $\pi^{\mu\nu}_{\ \ \kappa}$ of the metric, we see that the Lagrange multipliers $\Lambda$ must be equal to the momenta $b$:
\begin{equation}\label{Lambda=a}
    \Lambda^\kappa_{\mu\nu} = b^\kappa_{\mu\nu} \ .
\end{equation}
Consequently, we obtain the following gravitational field equations:
\begin{eqnarray}\label{P-jako-K}
  P_\lambda^{\mu\nu\kappa} &=& \frac {\partial L}{\partial
  K^\lambda_{\mu\nu\kappa }}\ , \\
  \nabla_\kappa  P_\lambda^{\mu\nu\kappa} &=& \frac {\partial L}{\partial
  \Gamma^\lambda_{\mu\nu}}  +\left( b^\mu_{\lambda\kappa} \pi^{\nu\kappa} +b^\nu_{\lambda\kappa} \pi^{\mu\kappa}
    - \delta^{(\mu}_\lambda b^{\nu )}_{\alpha\beta}\pi^{\alpha\beta}\right) \ , \label{a-rozw} \\
    \partial_\kappa  b^\kappa_{\mu\nu} &=& \frac {\partial L}{\partial
  \pi^{\mu\nu}} +\left( b^\kappa_{\lambda\mu} \Gamma^\lambda_{\nu\kappa} + b^\kappa_{\lambda\nu} \Gamma^\lambda_{\mu\kappa} - b^\kappa_{\mu\nu} \Gamma^\lambda_{\kappa\lambda}
    \right) \ , \label{a-nabla}
\end{eqnarray}
plus the standard Euler-Lagrange equations for the matter fields: $\partial_\kappa p^\kappa =  \frac {\partial L}{\partial \varphi}$, where $p^\kappa =  \frac {\partial L}{\partial \varphi_\kappa}$.

The last term in equation \eqref{a-nabla} combines, together with the left hand side, to the covariant derivative and we obtain:
\begin{equation}\label{a-nabla-1}
    \nabla_\kappa  b^\kappa_{\mu\nu} = \frac {\partial L}{\partial
  \pi^{\mu\nu}} \ .
\end{equation}
Equation \eqref{P-jako-K} expresses the momentum $P$ in terms of the curvature $K$. Finally, \eqref{a-rozw} can be easily solved with respect to the momenta $b^\kappa_{\mu\nu}$:
\begin{eqnarray*}\label{a-solved}
    b^\lambda_{\mu\nu} &=& \frac 12 \left\{ - \pi_{\mu\alpha}\delta^\lambda_\beta \delta^\sigma_\nu
    - \pi_{\nu\alpha}\delta^\lambda_\beta \delta^\sigma_\mu
    +\pi^{\lambda\sigma}\left( \pi_{\mu\alpha} \pi_{\nu\beta} - \frac 12 \pi_{\mu\nu} \pi_{\alpha\beta}
    \right)
    \right. \\
    &+& \left.  \frac 13 \pi_{\mu\nu}\delta^\sigma_\alpha \delta^\lambda_\beta \right\} \left(
    \nabla_\kappa  P_\sigma^{\alpha\beta\kappa} - \frac {\partial L}{\partial   \Gamma^\sigma_{\alpha\beta}}
  \right) \ ,
\end{eqnarray*}
where $\pi_{\mu\nu}$ is the inverse matrix of $\pi^{\mu\nu}$. Plugging this result into \eqref{a-nabla-1} we finally obtain gravitational field equations in the covariant form. They contain second order derivatives $\nabla_\lambda \nabla_\kappa  P_\sigma^{\alpha\beta\kappa}$ of the momenta, i.e. fourth order derivatives of the metric $\pi$.

\section{Proof of the Theorem}\label{proof}

We begin with generating formula \eqref{varia-A-G} and perform the Legendre transformation between the connection $\Gamma$ and the momentum $P$:
\begin{equation}\label{Leg-1}
    \partial_\kappa \left(
    P_\lambda^{\mu\nu\kappa} \delta \Gamma^\lambda_{\mu\nu} \right)
    =\delta \partial_\kappa \left(
    P_\lambda^{\mu\nu\kappa} \Gamma^\lambda_{\mu\nu} \right) -
    \partial_\kappa \left(\Gamma^\lambda_{\mu\nu} \delta
    P_\lambda^{\mu\nu\kappa}   \right) \ .
\end{equation}
We have
\begin{eqnarray*}
  \partial_\kappa \left(
    P_\lambda^{\mu\nu\kappa} \Gamma^\lambda_{\mu\nu} \right) &=& \left(\partial_\kappa
    P_\lambda^{\mu\nu\kappa} \right) \Gamma^\lambda_{\mu\nu} + P_\lambda^{\mu\nu\kappa}\Gamma^\lambda_{\mu\nu\kappa}\ .
\end{eqnarray*}
But:
\begin{eqnarray*}
  P_\lambda^{\mu\nu\kappa}\Gamma^\lambda_{\mu\nu\kappa} &=& P_\lambda^{\mu\nu\kappa}\left(\Gamma^\lambda_{\mu\nu\kappa} + \Gamma^{\lambda}_{\sigma\kappa}   \Gamma^{\sigma}_{\mu\nu} \right) - P_\lambda^{\mu\nu\kappa}\Gamma^{\lambda}_{\sigma\kappa}   \Gamma^{\sigma}_{\mu\nu}
  \\
    &=& P_\lambda^{\mu\nu\kappa} K^\lambda_{\mu\nu\kappa} - P_\lambda^{\mu\nu\kappa}\Gamma^{\lambda}_{\sigma\kappa}   \Gamma^{\sigma}_{\mu\nu} \ .
\end{eqnarray*}
Hence,
\begin{eqnarray}\label{H-dodatek}
  \partial_\kappa \left(
    P_\lambda^{\mu\nu\kappa} \Gamma^\lambda_{\mu\nu} \right) &=& \left(\partial_\kappa
    P_\lambda^{\mu\nu\kappa} \right) \Gamma^\lambda_{\mu\nu} - P_\lambda^{\mu\nu\kappa}\Gamma^{\lambda}_{\sigma\kappa}   \Gamma^{\sigma}_{\mu\nu} + P_\lambda^{\mu\nu\kappa} K^\lambda_{\mu\nu\kappa} \nonumber \\
    &=& \left(\nabla_\kappa
    P_\lambda^{\mu\nu\kappa} \right) \Gamma^\lambda_{\mu\nu} + P_\lambda^{\mu\nu\kappa}\Gamma^{\lambda}_{\sigma\kappa}   \Gamma^{\sigma}_{\mu\nu} + P_\lambda^{\mu\nu\kappa} K^\lambda_{\mu\nu\kappa}\ ,
\end{eqnarray}
the last equality being the consequence of identity \eqref{nabla-pi-konc}.
Putting the variation of this quantity on the left hand side we obtain from \eqref{varia-A-G}:
\begin{eqnarray}
    \delta \tilde{\Lambda} &=& \partial_\kappa \left( - \Gamma^\lambda_{\mu\nu} \delta P_\lambda^{\mu\nu\kappa} + b^\kappa_{\mu\nu} \delta \pi^{\mu\nu} + p^\kappa \delta \varphi\right) \ ,\label{varia-P}
\end{eqnarray}
where the new generating function
\begin{equation}\label{new-Lambda}
   \tilde{\Lambda} = L -\left(\nabla_\kappa
    P_\lambda^{\mu\nu\kappa} \right) \Gamma^\lambda_{\mu\nu} - P_\lambda^{\mu\nu\kappa}\Gamma^{\lambda}_{\sigma\kappa}   \Gamma^{\sigma}_{\mu\nu} - P_\lambda^{\mu\nu\kappa} K^\lambda_{\mu\nu\kappa}
\end{equation}
has to be expressed in terms of the ``control variables'', i.e.: $P$, $\pi$, $\varphi$ and their derivatives. Also the curvature tensor $K$ has to be expressed in terms of these. For this purpose we must solve field equations \eqref{P-jako-K} with respect to $K$.

Now, we use  decomposition \eqref{P-irreducible} of $P$. The antisymmetric part  ${\cal F}^{\mu\nu}$ drops out because $\Gamma$ is the metric connection . Hence, we have:
\begin{equation}\label{P-irreducible-sym}
    P_\lambda^{\mu\nu\kappa} =  \left(\delta_\lambda^\kappa \tilde{\pi}^{\mu\nu} - \delta_\lambda^{(\mu} \tilde{\pi}^{\nu)\kappa}  \right) + p_\lambda^{\mu\nu\kappa} \ ,
\end{equation}
and this is how the new metric $\tilde{\pi}^{\mu\nu}$ arises in the theory, downgrading the old one $\pi^{\mu\nu}$ to the level of matter fields. Similarly as in \eqref{pidGamma}, we have
\begin{equation}\label{p-Ga}
    \Gamma^\lambda_{\mu\nu} \delta P_\lambda^{\mu\nu\kappa} = A^\lambda_{\mu\nu} \delta \tilde{\pi}^{\mu\nu} +
    \Gamma^\lambda_{\mu\nu} \delta p_\lambda^{\mu\nu\kappa} \ ,
\end{equation}
where $A$ has been defined by formula \eqref{A-defi} in terms of the old connection $\Gamma^\lambda_{\mu\nu}$. Denoting by $\tilde{\Gamma}^\lambda_{\mu\nu}$ the new connection defined by the new metric $\tilde{\pi}^{\mu\nu}$ and by $\tilde{A}^\lambda_{\mu\nu}$ the corresponding objects built of its components, we can rewrite the generating formula \eqref{varia-P} in the following way:
\begin{eqnarray}
    \delta  \tilde{\Lambda} &=& \partial_\kappa \left( \left(- \tilde{A}^\lambda_{\mu\nu} + S^\lambda_{\mu\nu} \right)\delta \tilde{\pi}^{\mu\nu} - \Gamma^\lambda_{\mu\nu} \delta p_\lambda^{\mu\nu\kappa} + b^\kappa_{\mu\nu} \delta \pi^{\mu\nu} + p^\kappa \delta \varphi\right) \ ,\label{varia-P-pi}
\end{eqnarray}
where:
\[
    S^\lambda_{\mu\nu} := \tilde{A}^\lambda_{\mu\nu} - A^\lambda_{\mu\nu} \ .
\]
This is precisely formula \eqref{first-symp-2} for the new metric $\tilde{\pi}^{\mu\nu}$ interacting with three different groups of matter fields: 1)
the original matter field $\varphi$ (together with its momentum $p^\kappa$), 2) the old metric $\pi^{\mu\nu}$ (together with its momentum $b^\kappa_{\mu\nu}$) and 3) the field $p_\lambda^{\mu\nu\kappa}$ (together with its momentum\footnote{To better understand the symplectic structure of the theory it is useful to rewrite the term containing $p$ in \eqref{varia-P-pi}:
\[
 \Gamma^\lambda_{\mu\nu} \delta p_\lambda^{\mu\nu\kappa} = G^{\kappa\lambda}_{\mu\nu\sigma} \delta p_\lambda^{\mu\nu\sigma} \ ,
\]
where the momentum $G^{\kappa\lambda}_{\mu\nu\sigma}$ has been built from $\delta^\kappa_\sigma\Gamma^\lambda_{\mu\nu}$ by projecting it on the subspace of tensors fulfilling all the algebraic symmetries of $p_\lambda^{\mu\nu\sigma}$. Hence, momentum canonically conjugate to $p$ is represented by $G$, which carries precisely the same information as $\Gamma$ does.}, which is equal to $- \Gamma^\lambda_{\mu\nu}$). Hence, the new matter fields $\phi$ which appear in the thesis of the Theorem represent the second and the last ones of this triple: $\phi =(\pi^{\mu\nu} , p_\lambda^{\mu\nu\kappa})$. This observation completes the proof of the Theorem.

To describe better the structure of $\tilde{\Lambda}$ we use decomposition \eqref{K-irreducible} of the curvature tensor:
\begin{equation}\label{K-irreducible-sym}
    K^\lambda_{\mu\nu\kappa} = - \frac 19 \left( \delta^\lambda_\mu K_{\nu\kappa} +
    \delta^\lambda_\nu K_{\mu\kappa} - 2 \delta^\lambda_\kappa K_{\mu\nu} \right) +
     U^\lambda_{\mu\nu\kappa} \ ,
\end{equation}
(antisymmetric part $F_{\mu\nu}$ of the Ricci vanishes identically because $\Gamma$ is metric). Identity \eqref{P-irreducible-0} implies that the definition \eqref{P-jako-K} of $P_\lambda^{\mu\nu\kappa}$ splits into two independent components:
\begin{eqnarray}
   \tilde{\pi}^{\mu\nu} &=& \frac {\partial L}{\partial
  K_{\mu\nu}}\ , \label{K-row}\\
   p_\lambda^{\mu\nu\kappa} &=& \frac {\partial L}{\partial
  U^\lambda_{\mu\nu\kappa }}\ .\label{U-row}
\end{eqnarray}
Moreover, due to \eqref{P-irreducible-sym}, we have:
\begin{eqnarray*}
  P_\lambda^{\mu\nu\kappa} K^\lambda_{\mu\nu\kappa} &=& \tilde{\pi}^{\mu\nu} K_{\mu\nu}
    + p_\lambda^{\mu\nu\kappa} U^\lambda_{\mu\nu\kappa} \ , \\
  \left(\nabla_\kappa
    P_\lambda^{\mu\nu\kappa} \right) \Gamma^\lambda_{\mu\nu} &=&
    \left(\nabla_\lambda\tilde{\pi}^{\mu\nu}\right) A^\lambda_{\mu\nu} +
    \left(\nabla_\kappa
    p_\lambda^{\mu\nu\kappa} \right) \Gamma^\lambda_{\mu\nu} \ ,  \\
    P_\lambda^{\mu\nu\kappa}\Gamma^{\lambda}_{\sigma\kappa}   \Gamma^{\sigma}_{\mu\nu}
    &=& \tilde{\pi}^{\mu\nu} \left(\Gamma^{\sigma}_{\mu\nu} \Gamma^{\kappa}_{\sigma\kappa}
    - \Gamma^{\sigma}_{\kappa\mu} \Gamma^{\kappa}_{\sigma\nu}
    \right) +
    p_\lambda^{\mu\nu\kappa}\Gamma^{\lambda}_{\sigma\kappa}   \Gamma^{\sigma}_{\mu\nu} \\
     \left(\nabla_\kappa
    p_\lambda^{\mu\nu\kappa} \right) \Gamma^\lambda_{\mu\nu} + p_\lambda^{\mu\nu\kappa}\Gamma^{\lambda}_{\sigma\kappa}   \Gamma^{\sigma}_{\mu\nu}
    &=& \left(\partial_\kappa
    p_\lambda^{\mu\nu\kappa} \right) \Gamma^\lambda_{\mu\nu} - p_\lambda^{\mu\nu\kappa}\Gamma^{\lambda}_{\sigma\kappa}   \Gamma^{\sigma}_{\mu\nu}
    \ .
\end{eqnarray*}
Hence, formula \eqref{L-from-Lambda} implies the following form of the new matter Lagrangian:
\begin{eqnarray}
    \tilde{L}_{Matter} &=& \tilde{L}_{Matter}\left( \tilde{\pi}, \partial \tilde{\pi}; \varphi , \partial \varphi , \pi , \partial \pi , p , \partial p
    \right)  \\
    &=&\tilde{\Lambda} - \tilde{\pi}^{\mu\nu} \left(
\tilde{\Gamma}^{\lambda}_{\mu\sigma} \tilde{\Gamma}^{\sigma}_{\nu\lambda} -
\tilde{\Gamma}^{\lambda}_{\mu\nu} \tilde{\Gamma}_{\lambda\sigma}^\sigma
    \right)  \nonumber \\
    &=& L - \tilde{\pi}^{\mu\nu} K_{\mu\nu}
    - p_\lambda^{\mu\nu\kappa} U^\lambda_{\mu\nu\kappa}
    \nonumber \\
    &-&\left(\nabla_\lambda\tilde{\pi}^{\mu\nu}\right) A^\lambda_{\mu\nu} -
    \left(\partial_\kappa
    p_\lambda^{\mu\nu\kappa} \right) \Gamma^\lambda_{\mu\nu} +
    p_\lambda^{\mu\nu\kappa}\Gamma^{\lambda}_{\sigma\kappa}   \Gamma^{\sigma}_{\mu\nu} \nonumber \\
    &-& \tilde{\pi}^{\mu\nu} \left(\Gamma^{\lambda}_{\mu\sigma} \Gamma^{\sigma}_{\nu\lambda}
    - \Gamma^{\lambda}_{\mu\nu} \Gamma^{\sigma}_{\lambda\sigma}
    \right)
    -\tilde{\pi}^{\mu\nu} \left(
\tilde{\Gamma}^{\lambda}_{\mu\sigma} \tilde{\Gamma}^{\sigma}_{\nu\lambda} -
\tilde{\Gamma}^{\lambda}_{\mu\nu} \tilde{\Gamma}_{\lambda\sigma}^\sigma
    \right)
    \ . \label{L-matter-tilde}
\end{eqnarray}
Here, $U^\lambda_{\mu\nu\kappa}$ and $K_{\mu\nu}$ have to be expressed in terms of the remaining variables. For this purpose  equations \eqref{K-row} and \eqref{U-row} must be solved with respect to $K$ and $U$.

{\bf Remark 1:} If the original Lagrangian $L$ does not depend upon the Weyl tensor $U^\lambda_{\mu\nu\kappa}$, but only upon the Ricci $K_{\mu\nu}$, then \eqref{U-row} implies $p_\lambda^{\mu\nu\kappa}\equiv 0$ and the new matter variables reduce to the ``old metric'' $\pi$. This case was thoroughly analyzed in \cite{Jakubiec}. In particular, the Sacharov theory \cite{Sacharov} (see also \cite{Ferraris}) belongs to this class.

{\bf Remark 2:} Another interesting class of theories is given by Lagrangians having the following structure:
\begin{equation}\label{class-Weyl}
    L = L_{Hilbert} + L_{Weyl} \ ,
\end{equation}
where $L_{Hilbert}$ is given by \eqref{Hilb} (i.e.~we have $L_{Hilbert}=\pi^{\mu\nu} K_{\mu\nu}$) and where $L_{Weyl}$ depends upon the Weyl tensor and the metric. Equation \eqref{K-row} implies $\tilde{\pi}^{\mu\nu} = \pi^{\mu\nu}$.
We have, therefore $\tilde{\Gamma}^{\lambda}_{\mu\sigma}  =
{\Gamma}^{\lambda}_{\mu\sigma} $ and, consequently, $\nabla_\lambda\tilde{\pi}^{\mu\nu} \equiv 0$.
Hence, the only matter field of the theory is $p$ given by \eqref{U-row}:
\begin{equation}\label{class-p}
    p_\lambda^{\mu\nu\kappa} = \frac {\partial L}{\partial
  U^\lambda_{\mu\nu\kappa }}\ .
\end{equation}

{\bf Example:} Consider the following Lagrangian:
\begin{eqnarray}
    L &=& \frac 1{16\pi} \sqrt{|g|} R +  \frac c2 {\sqrt{|g|}} U^\lambda_{\mu\nu\kappa} U^\sigma_{\alpha\beta\gamma} g_{\lambda\sigma} g^{\mu\alpha} g^{\nu\beta} g^{\kappa\sigma} \nonumber \\
    &=& \pi^{\mu\nu} K_{\mu\nu} + \frac c2 \sqrt{|g|} \left( U \cdot U \right) \ , \label{Weyl}
\end{eqnarray}
where the standard Hilbert term has been appended by the square of the Weyl tensor multiplied by an arbitrary constant $c$.
Equations \eqref{K-row} and \eqref{U-row} imply:
\begin{eqnarray}
   \tilde{\pi}^{\mu\nu} &=& \pi^{\mu\nu}\ , \label{K-row-ex}\\
   p_\lambda^{\mu\nu\kappa} &=& c \sqrt{|g|}U^\sigma_{\alpha\beta\gamma} g_{\lambda\sigma} g^{\mu\alpha} g^{\nu\beta} g^{\kappa\sigma} \ .\label{U-row-ex}
\end{eqnarray}
We have, therefore,
\begin{equation}\label{L-Kpi}
    L - \tilde{\pi}^{\mu\nu} K_{\mu\nu}
    - p_\lambda^{\mu\nu\kappa} U^\lambda_{\mu\nu\kappa} =
    - \frac 1{2 c \sqrt{|g|}}  \
    p_\lambda^{\mu\nu\kappa} p_\sigma^{\alpha\beta\gamma} g^{\lambda\sigma} g_{\mu\alpha} g_{\nu\beta} g_{\kappa\sigma} \ .
\end{equation}
Hence, the matter Lagrangian \eqref{L-matter-tilde} for the field $p$ reduces to:
\begin{equation}\label{New-L-ex}
    \tilde{L}_{Matter} =
    - \frac 1{2 c \sqrt{|g|}}  \
    p_\lambda^{\mu\nu\kappa} p_\sigma^{\alpha\beta\gamma} g^{\lambda\sigma} g_{\mu\alpha} g_{\nu\beta} g_{\kappa\sigma}
    - \left(\partial_\kappa
    p_\lambda^{\mu\nu\kappa} \right) \Gamma^\lambda_{\mu\nu} +
    p_\lambda^{\mu\nu\kappa}\Gamma^{\lambda}_{\sigma\kappa}   \Gamma^{\sigma}_{\mu\nu} \ .
\end{equation}

\section{Conclusions}
The main advantage of the reformulation presented above is the applicability of the standard Hamiltonian formalism developed for purposes of general relativity theory. In particular, the ``positive energy'' theorem applies here if and only if the matter energy is positive. This is probably the simplest way to analyze stability of different models of this type.

There might be doubts about which one of the two metric tensors is ``the true one''. This question was already considered by Higgs, who had the following remark: {\em
it seems likely that more direct physical significance may be attached to the new metric than to the original dynamic variables $g$, which enter into the action principle} (see \cite{Higgs}).

In this context I want to stress that already in the absolutely standard Einstein formulation the gravitational waves can propagate along different ``light cones'' than the ones defined by the metric tensor. Indeed, if the matter Lagrangian contains connection coefficients (necessary, e.g., for the covariant derivatives of the matter fields) then the energy momentum tensor contains also second derivatives of the metric, contained in the last term of equation \eqref{Einst-cone}). Hence, expression $G_{\mu\nu} - 8\pi T_{\mu\nu}$ contains second derivatives multiplied not only by the metric tensor (coming from $G_{\mu\nu}$) but also by the functions of the matter fields (coming from $T_{\mu\nu}$). The effective light cone is, therefore, different from the one defined by the metric. The existence of two metric tensors is, therefore, not so controversial as one could feel at the beginning.

Having already accepted the existence of different metric tensors in the theory, the question: ``which one among them is more physical than the remaining ones'' is irrelevant as far as the dynamical properties of the field evolution are considered. Indeed, these properties do not depend upon the set of equivalent variables in \eqref{change}, which we use to parameterize field configurations. But the very mathematical structure of the theory distinguishes our metric $\tilde{\pi}^{\mu\nu}$, arising as a momentum canonically conjugate to the Ricci tensor $K_{\mu\nu}$. Using it, the gravitational part of field equations will always be written in the universal form of Einstein equations $G^{\mu\nu}(\tilde{g}) = 8\pi \tilde{T}^{\mu\nu}$, no matter how exotic and complicated is the Lagrangian \eqref{inv} of the theory.

Of course, it is hard to believe that the theory like \eqref{New-L-ex} has any fundamental value. In my opinion it can be treated as merely a phenomenological theory. Nevertheless, our theory shows that, instead of ``generalizing'' general relativity theory, one can concentrate on inventing new matter fields describing phenomena which we want to model (e.g.~black energy). In this context our theorem can be a good starting point. In particular, the ``purely affine'' theory \eqref{Lagr-aff}, which does not contain any ``primary metric'' $\pi$, is especially interesting. Here, there is a unique metric tensor $\tilde{\pi}$, arising dynamically as a momentum canonically conjugate to the Ricci part of the curvature.

Using ideas presented in \cite{Kij-Moreno}, one can easily generalize our Theorem to the case of Lagrangians depending not only upon curvature tensor, but also upon its (covariant) derivatives. In this case matter field equations will be of higher differential order, whereas gravitational field will be always described by the conventional Einstein equations. The only difference would be the dependence of the matter energy-momentum tensor \eqref{var-varphi} upon higher derivatives of the matter fields. Consequently, the positivity of the total mass (and the stability of the theory) can be analyzed in terms of the conventional tools of Hamiltonian gravity.

\section*{Acknowledgments}

This research was supported by Narodowe Centrum Nauki, Poland (grant DEC-2011/03/B/ST1/02625). The author is very much indebted to Marek Demia\'nski for discussions concerning ``generalizations of general relativity theory''. Many thanks are also due to Leszek Soko{\l}owski for discussions concerning Lagrangians depending upon the Weyl tensor. These discussions provided the main inspirations to put together my observations concerning metric-affine variational principles and to prepare this article.

\appendix

\section*{Appendix: Proof of the formula \eqref{nabla-pi-konc}}

\begin{eqnarray*}
    \nabla_\kappa P_\lambda^{\mu\nu\kappa}
    &=& \partial_\kappa P_\lambda^{\mu\nu\kappa} -
    P_\sigma^{\mu\nu\kappa} \Gamma^\sigma_{\lambda\kappa}
    + P_\lambda^{\sigma\nu\kappa} \Gamma^{\mu}_{\sigma\kappa}
    + P_\lambda^{\mu\sigma\kappa} \Gamma^{\nu}_{\sigma\kappa}
    + P_\lambda^{\mu\nu\sigma} \Gamma^{\kappa}_{\sigma\kappa}
    - P_\lambda^{\mu\nu\kappa} \Gamma^{\sigma}_{\kappa\sigma} \\
    &=& \partial_\kappa P_\lambda^{\mu\nu\kappa} -
    P_\sigma^{\mu\nu\kappa} \Gamma^\sigma_{\lambda\kappa}
    + P_\lambda^{\sigma\nu\kappa} \Gamma^{\mu}_{\sigma\kappa}
    + P_\lambda^{\mu\sigma\kappa} \Gamma^{\nu}_{\sigma\kappa} \\
    &=&\partial_\kappa P_\lambda^{\mu\nu\kappa} -
    P_\sigma^{\mu\nu\kappa} \Gamma^\sigma_{\lambda\kappa}
    +\frac 12 \left( P_\lambda^{\sigma\nu\kappa} +P_\lambda^{\kappa\nu\sigma}
    \right) \Gamma^{\mu}_{\sigma\kappa}
    + \frac 12 \left( P_\lambda^{\mu\sigma\kappa} + P_\lambda^{\mu\kappa\sigma}
    \right)\Gamma^{\nu}_{\sigma\kappa} \ .
\end{eqnarray*}
But, as a consequence of identities \eqref{P-symmetries}, we have:
\[
  P_\lambda^{\sigma\nu\kappa} +P_\lambda^{\kappa\nu\sigma} = - P_\lambda^{\kappa\sigma\nu} \ ,
 \]
and
\[
    P_\lambda^{\mu\sigma\kappa} + P_\lambda^{\mu\kappa\sigma} = - P_\lambda^{\sigma\kappa\mu} \ .
\]
Plugging these into the previous equation we obtain \eqref{nabla-pi-konc}.


\begin{thebibliography}{0}


\bibitem{Stelle}
P.~Havas; General Relativity and Gravitation, 8 (1977) 631;\\
G.T.Horowitz and R.M.Wald;
Phys. Rev. D 17 (1978) 414;\\
K.S.Stelle; Gen. Relativ. Gravit. 9 (1978) 353;\\
K.I.Macrae and R.J.Rieger; Phys. Rev. D24 (1981) 2555;\\
A.Frenkel and  K.Brecher,ibid. 26 (1982) 368;\\
V.M\"uller and H.-J.Schmidt, Gen. Relativ. Gravit. 17 (1985) 769 and 971.

\bibitem{Higgs} G.~Stephenson; Il Nuovo Cimento, 9 (1958) 263;
P.~W.~Higgs; Il Nuovo Cimento, 11 (1959) 817.


\bibitem{Jakubiec}
A. Jakubiec and J. Kijowski:
Phys. Rev. D. 37 (1988) 1406;
Journ. Math. Phys. 30 (1989) 2923;
Journ. Math. Phys. 30 (1989) 1073.



\bibitem{Sacharov}
A.D.Sakharov, Dok. Akad. Nauk SSSR 177 (1967) 70;

\bibitem{Ferraris}
M.Ferraris, in {\em Atti del VI Convegno Nazionale di Relativita' Generale e Fisica della Gravitazione},
Firenze, 1984 (Pitagora, Bologna, Italy, 1986), p. 127;\\
R. Kerner, Gen. Relativ. Gravit. 14 (1982) 453.

\bibitem{FRA}
T.~P.~Sotiriou, S.~Liberati, J. Phys. Conf. Ser. 68 (2007) 012022,\\
S.~Capozziello, M.~De Laurentis, M.~Francaviglia,
S.~Mercadante, Found. Phys.  39 (2009) 1161. \\
S.~Capozziello, M.~De Laurentis, {\em Extended Theories of Gravity}, Physics Reports (2011) Elsevier

\bibitem{Jakubiec-Cauchy}
A.~Jakubiec, J. Kijowski~,
Journ. Math. Phys. 30 (1989) p. 2923 -- 2924.

\bibitem{Fock} V.~Fock {The theory of space time and gravitation} (translated from the Russian), Pergamon Press, London (1959)


\bibitem{WMT}
C.W. Misner, K.S. Thorne, J.A. Wheeler,
{\em Gravitation}, N.H. Freeman and Co, San Francisco, Cal. (1973).

\bibitem{CJK}
P. Chrusciel, J. Jezierski and J. Kijowski
{\em Hamiltonian Field Theory in the Radiating Regime}
monograph (174 pages), volume 70 of the series: Springer Lecture Notes in Physics, Monographs (2001);\\
P. Chrusciel, J. Jezierski, J. Kijowski
Phys. Rev. D 87 (2013)  124015

\bibitem{pieszy}

J.~Kijowski,
Gen. Relat. Grav.  {\bf 29} (1997) 307.

\bibitem{affine}

J.~Kijowski,
Gen. Relat. Grav. {\bf 9} (1978) 857;\\
J. Kijowski, R. Werpachowski,
Rep. Math. Phys. 59 (2007) 1.

\bibitem{Tulcz-Kij}

J. Kijowski and W.M. Tulczyjew, {\it A
Symplectic Framework for Field Theories}, Lecture Notes in Physics
No.107 (Springer-Verlag, Berlin, 1979)

\bibitem{Wald} S. R. Green, J. S. Schiffrin and R. M. Wald,
arXiv:1309.0177v2 [gr-qc]


\bibitem{Kij-Moreno}
J.~Kijowski and G.~Moreno,
{\em Symplectic structures related with higher order variational problems},
Int. Journ. Geom. Meth. Modern Phys., 12 (2015) 1550084


\end{thebibliography}
\end{document}